\documentclass[english,onecolumn,aps,superscriptaddress,amsmath,amssymb,12pt,showkeys,prb]{revtex4-1}
\usepackage[utf8]{inputenc}
\usepackage{graphicx}
\usepackage{subfigure}
\usepackage{dcolumn}
\usepackage{bm} 
\usepackage[normalem]{ulem}
\usepackage[pdftex,usenames,dvipsnames]{color}
\usepackage{hyperref}
\usepackage{soul}

\allowdisplaybreaks

\makeatletter
\newcommand*{\supplementary}{%
  \close@column@grid
  \clearpage
  \twocolumngrid
}
\makeatother

\def\eqref#1{(\ref{#1})}

\definecolor{tangerine}{rgb}{0.944,0.522,0}
\definecolor{verde}{rgb}{0.,0.6,0}
\definecolor{rosso}{rgb}{0.9,0.0,0.2}
\definecolor{blue}{rgb}{0.0,0.0,0.8}

\newcommand{\editor}[2]{%
  \expandafter\newcommand\csname #1note\endcsname[1]{%
    \textcolor{#2}{(\textbf{#1:} \textit{\small ##1})}}%
  \expandafter\newcommand\csname #1\endcsname[1]{%
    \textcolor{#2}{##1}}%
  \expandafter\newcommand\csname #1cancel\endcsname[1]{%
    \textcolor{#2}{\sout{##1}}}%
  \expandafter\newcommand\csname #1change\endcsname[2]{%
    \textcolor{#2}{\sout{##1} ##2}}%
  \newenvironment{#1text}{\color{#2}}{\color{black}}
}

\editor{AM}{tangerine}
\editor{FZ}{blue}
\editor{TB}{verde}
\editor{TL}{rosso}

\begin{document}
\title{Simulating diffusion properties of solid-state electrolytes via a neural network potential: Performance and training scheme}

\author{Aris Marcolongo}
\thanks{These authors contributed equally to the work.}
\affiliation{National Centre for Computational Design and Discovery of Novel Materials MARVEL, Cognitive Computing and Computational Sciences Department, IBM Research – Z\"{u}rich, Säumerstrasse 4, CH-8803 Rüschlikon, Switzerland}
\author{Tobias Binninger}
\thanks{These authors contributed equally to the work.}
\affiliation{National Centre for Computational Design and Discovery of Novel Materials MARVEL, Cognitive Computing and Computational Sciences Department, IBM Research – Z\"{u}rich, Säumerstrasse 4, CH-8803 Rüschlikon, Switzerland}
\author{Federico Zipoli}
\thanks{These authors contributed equally to the work.}
\affiliation{National Centre for Computational Design and Discovery of Novel Materials MARVEL, Cognitive Computing and Computational Sciences Department, IBM Research – Z\"{u}rich, Säumerstrasse 4, CH-8803 Rüschlikon, Switzerland}
\author{Teodoro Laino}
\affiliation{National Centre for Computational Design and Discovery of Novel Materials MARVEL, Cognitive Computing and Computational Sciences Department, IBM Research – Z\"{u}rich, Säumerstrasse 4, CH-8803 Rüschlikon, Switzerland}
\date{\today}
\begin{abstract}
The recently published DeePMD model, based on a deep neural network architecture, \cite{PhysRevLett.120.143001} brings the hope of solving the time-scale issue which often prevents the application of first principle molecular dynamics to physical systems. With this contribution we assess the performance of the DeePMD potential on a real-life application and model diffusion of ions in solid-state electrolytes. We consider as test cases the well known Li$_{10}$GeP$_2$S$_{12}$, Li$_7$La$_3$Zr$_2$O$_{12}$ and Na$_{3}$Zr$_2$Si$_2$PO$_{12}$. We develop and test a training protocol suitable for the computation of diffusion coefficients, which is one of the key properties to be optimized for battery applications, and we find good agreement with previous computations. Our results show that the DeePMD model may be a successful component of a framework to identify novel solid-state electrolytes.
\end{abstract}

\maketitle

\section{Introduction}
Optimization of solid-state electrolytes (SSE) can improve significantly the performance of all-solid-state batteries, enabling new technologies which can have a considerable impact on  our society by providing e.g. safer portable devices and electric vehicles with a longer battery life  \cite{kim_review_2015}.The importance of this topic is exemplified by the huge experimental and theoretical effort ongoing both in the industrial and academic environment. \\ 
An optimal candidate SSE needs to be stable with respect to both electrodes,  electronically insulating, and permit an easy flow of ionized cations from the anode to the cathode and vice versa \cite{weppner_engineering_2003,tikekar_design_2016}. The latter property can be computationally predicted via  the ionic diffusion coefficient from accurate first principle-molecular dynamics (FPMD) simulations \cite{Frenkel:1996:UMS:547952}. The screening of material databases is impeded by the computational cost required to arrive at reliable predictions. Therefore, computational groups have been devoting their efforts either to the search of descriptors \cite{muy_high-throughput_2019,d.sendek_holistic_2017,Gao2014,wang_design_2015}, or to the development of computational strategies to speed-up the simulations \cite{kahle_modeling_2018}.\\
In this context, machine learning models based for example on deep neural networks \cite{Goodfellow:2016:DL:3086952} have been developed to model the potential energy surface of molecules,  materials and liquids \cite{PhysRevLett.98.146401,PhysRevLett.104.136403,PhysRevLett.114.096405,botu_094306,doi:10.1021/acs.jpclett.6b01448,C6SC05720A}, offering the possibility to perform molecular dynamics (MD) simulations at a cost comparable to empirical classical force fields and with the accuracy of first-principles molecular dynamics at density functional theory (DFT) level. The ultimate goal can be seen in the development of a transferable universal model able to accurately predict the ionic diffusion properties of candidate SSE materials after being trained on a set of known materials. 
The recently developed open-source DeePMD model, based on a deep neural network (DNN) architecture \cite{PhysRevLett.120.143001,2018CoPhC.228..178W}, has already been tested for its transferability properties in Al--Mg systems \cite{PhysRevMaterials.3.023804}, and it is therefore a good candidate model for such a purpose. In this work, we develop training strategies and test the power of the DeePMD model for the prediction of diffusion properties of three well-known SSEs: Li$_{10}$GeP$_2$S$_{12}$ \ \text{(LGPS)} \cite{kamaya_lithium_2011,mo_first_2012,marcolongo_ionic_2017},  Li$_7$La$_3$Zr$_2$O$_{12}$ \ \text{(LLZO)}\cite{doi:10.1002/anie.200701144,doi:10.1021/jp5002463,PhysRevMaterials.3.035403,jalem_concerted_2013}, and Na$_{3}$Zr$_2$Si$_2$PO$_{12}$ \ \text{(NASICON)}\cite{BOILOT1988160,NASICON_2} . Overall, the deepMD model is able to reproduce the superionic behavior of SSEs and we reach with our training protocol an uncertainty on the computed diffusion coefficients of about 20\% or smaller, which is precise enough for applications. The activation energies computed from the DeepMD model are in line with previous FPMD simulations, when available, and compatible with experiments.

\section{Training protocol}
\begin{figure}
\includegraphics[trim=1 150 1 150, clip,width=\hsize]{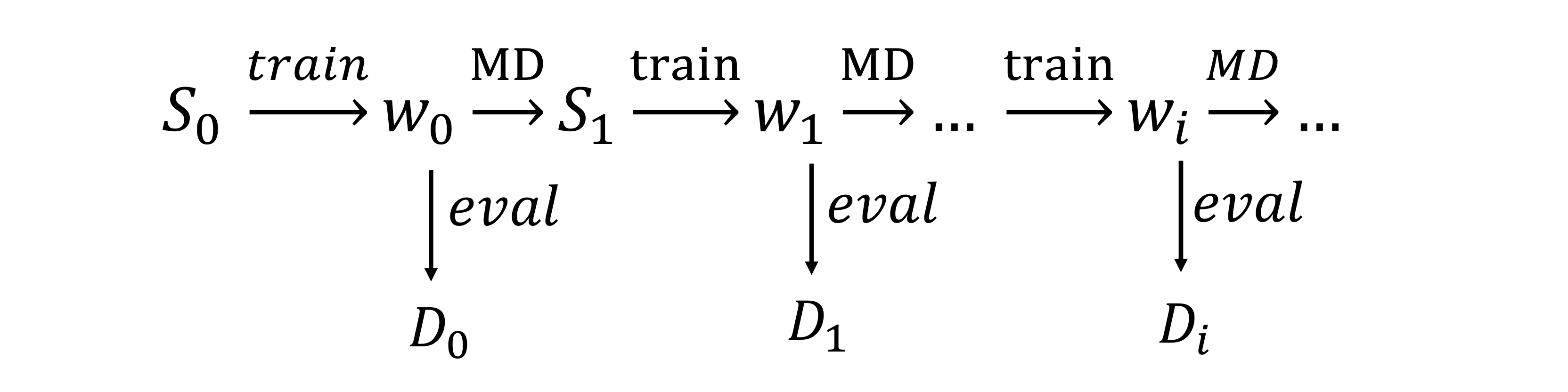}
\caption{The workflow defining the training protocol used in this work. A starting training set of configurations $S_0$ is chosen. A \emph{training step} consists in a reoptimization of the parameters and the creation of an enlarged training set via a MD simulation: $S_i \xrightarrow[]{\text train} w_i \xrightarrow[]{\text{MD}} S_{i+1}$. The temperature for the MD steps rises from $T_{min}$ to $T_{max}$ and all training steps belonging to one temperature ramp constitute a \emph{training loop}. Convergence is controlled via the evolution of the diffusion coefficient computed in the evaluation step, until stationarity is achieved.}
\label{fig::loop}
\end{figure}

We consider a generic model of a potential energy surface with parameters $w$ which, in the case of a DNN, are the connections between the neurons, also called \emph{weights} \cite{Goodfellow:2016:DL:3086952}. Let $\Omega$ be the phase space of the system, whose configurations are defined by the coordinates and velocities of all atoms $\{ q,p \} \in \Omega$. Any fitting procedure starts by defining a loss function for a given configuration. The most common choice is the squared loss:
\begin{equation}
\mathcal{L}_{w}(\{ q,p \}) \equiv |\mathcal{O}_{w}(\{ q,p \})-\mathcal{O}_{exact}(\{ q,p \})|^2,
\end{equation}
where $\mathcal{O}: \Omega \rightarrow \mathbb{R}^N$ is a selected set of (real) observables, in general forces, energy and virial. The parameters $w$ are then fitted to minimize the loss function $\mathcal{L}_{\omega} \equiv \langle \mathcal{L}_{\omega}(\{ q,p \}) \rangle_{D}$ averaged over a fixed distribution $D$ on $\Omega$. \\
For molecular dynamics simulations, the natural choice of $D$ is the canonical Boltzmann distribution $P_{B}(\{ q,p \}) \sim \exp[-U(\{ q,p \})/{k_BT}]$, which provides the probability of each configuration, but in order to sample this distribution one would need to perform beforehand expensive molecular dynamics or Monte Carlo simulations. We must therefore resort to a distribution defined by a training set $S \in \Omega$. When training DNNs, a bad choice of $S$ can lead to overfitting and poor generalization.\\
The authors of the DeePMD model recently introduced the concept of active learning to
circumvent these problematics \cite{PhysRevMaterials.3.023804}. A starting set of configurations $S_0$ is chosen, which serves
to train a first approximate model. The initial training set is then augmented in an iterative
procedure that is divided into three conceptual steps: Exploration, Labeling and Training,
combined in an iterative way. In the exploration phase, a subset of the phase space is
sampled via molecular dynamics using an approximate model. In order to decide whether to
add a new configuration to the training set S, the authors proposed to fit different models
changing the initialization of the weights. If the fitted models disagree on the predicted en-
ergy and forces, the configuration is labelled, i.e. the corresponding exact energy and forces
are computed via DFT, and added to the training set. When the training set is sufficiently
augmented, an improved model is trained.\\
In this work, we follow the same conceptual steps of Ref. \cite{PhysRevMaterials.3.023804} and we develop a learning strategy where the training set is augmented over time. However, we simplify the labelling step and aim at a Boltzmann distributed training set via a self-consistent procedure. Our general strategy is summarized in Fig. \ref{fig::loop}. We start from an initial set of configurations $S_0$ and a set of parameters $w_0$. For the exploration phase a molecular dynamics simulation is performed using the approximate model. We then sample the approximate canonical distribution $P_{B}^{model} \sim \exp[-U^{model}(\{ q,p \})/{k_BT}]$ by selecting equidistant snapshots from the MD trajectory performed with a Nos\'e-Hoover thermostat \cite{PhysRevA.31.1695}. All selected configurations are added to the training set, the DNN weights are retrained and the procedure continues. We call \emph{training step} a retraining of the model followed by an update of the training set. For this work, we chose the MD temperature at which the additional training configurations are sampled in each training step to raise from $300$\,K to $1000$\,K, in steps of $50$\,K, and we call one such temperature ramp a \emph{training loop}. A complete training for one material can require several loops. We control the convergence of the training workflow by introducing an \emph{evaluation} step where the quantity of interest, in our case the diffusion coefficient, is determined. Ideally, when convergence is obtained, the DFT energies and forces are consistent with the ones predicted by the model for all configurations sampled. Otherwise, they are added to the training set until self-consistency is achieved and the diffusion coefficient predicted in the evaluation step becomes stationary.\\
While similar, the approach of Ref. \cite{PhysRevMaterials.3.023804} focuses its convergence criterion on an estimate of the uncertainty of the underlying potential energy model. Our criterion focuses on an explicit check of the quality of the fitted model on a Boltzmann distributed set in order to avoid overfitting as much as possible. \\ 
We test and compare two different choices for the starting distribution $S_0$:
\begin{itemize}
\item We perform short FPMD simulations with constant number of particles, volume and temperature (NVT) at three different temperatures ($300, 600$ and $900$\,K), using the experimental volume provided in the material cif-file.
\item We sample $S_0$ from long MD simulations using a polarizable force field (PFF) that was preliminary fitted using the procedure in Ref. \cite{fzi}. PFF-MD simulations are performed at the same three different temperatures, but the volume is equilibrated beforehand via PFF-MD simulations at constant number of particles, pressure and temperature (NPT). Here, a larger number of atoms can be used for producing the starting training set $S_0$.
\end{itemize}
The first procedure is straightforward using available FPMD codes, whereas the second one requires the expertise to fit the coefficients of semi-empirical models. By comparing the results of the training loop with different choices of $S_0$, we obtain information on the effect of the different meta-parameters on the final result. Further details are reported in appendix A.

\begin{figure}
\includegraphics[width=\hsize]{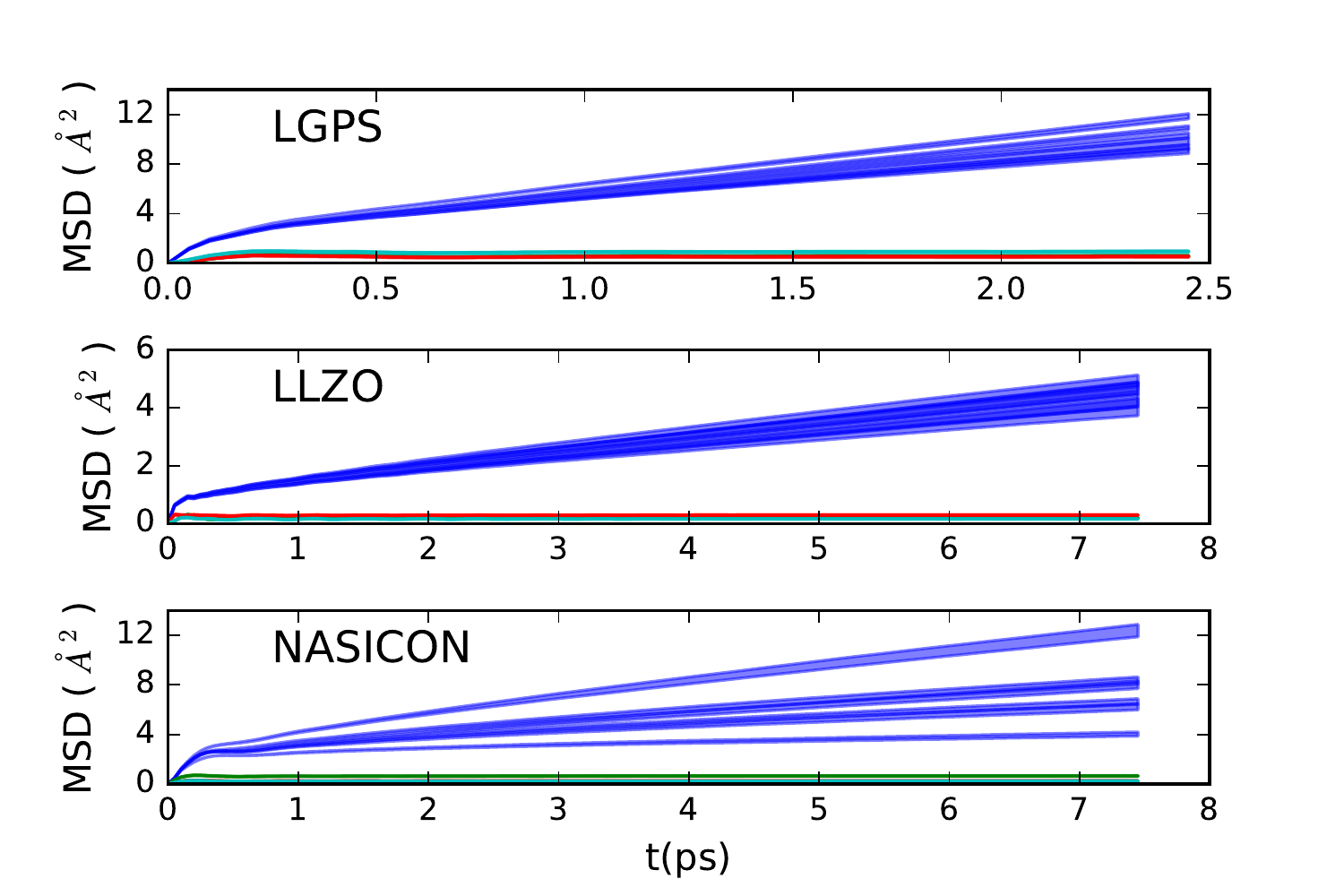}
\caption{ Species resolved MSDs obtained from NVE MD simulations with the DeePMD model, at a target temperature of $625$\,K. Different curves refer to models evaluated in different training steps of a second iteration of the training loop. $S_0$ was sampled from a short FPMD trajectory for each material. The ion conduction behavior of the SSEs is confirmed, with a linear Lithium/Sodium MSD reported in blue with an associated one $\sigma$ confidence interval. Only for NASICON, we find a significant spread of the MSDs of the carriers. 
We report with different colors the bounded MSDs of the non-diffusive species.}
\label{fig::all_msd}
\end{figure}
\begin{figure}
\includegraphics[width=\hsize]{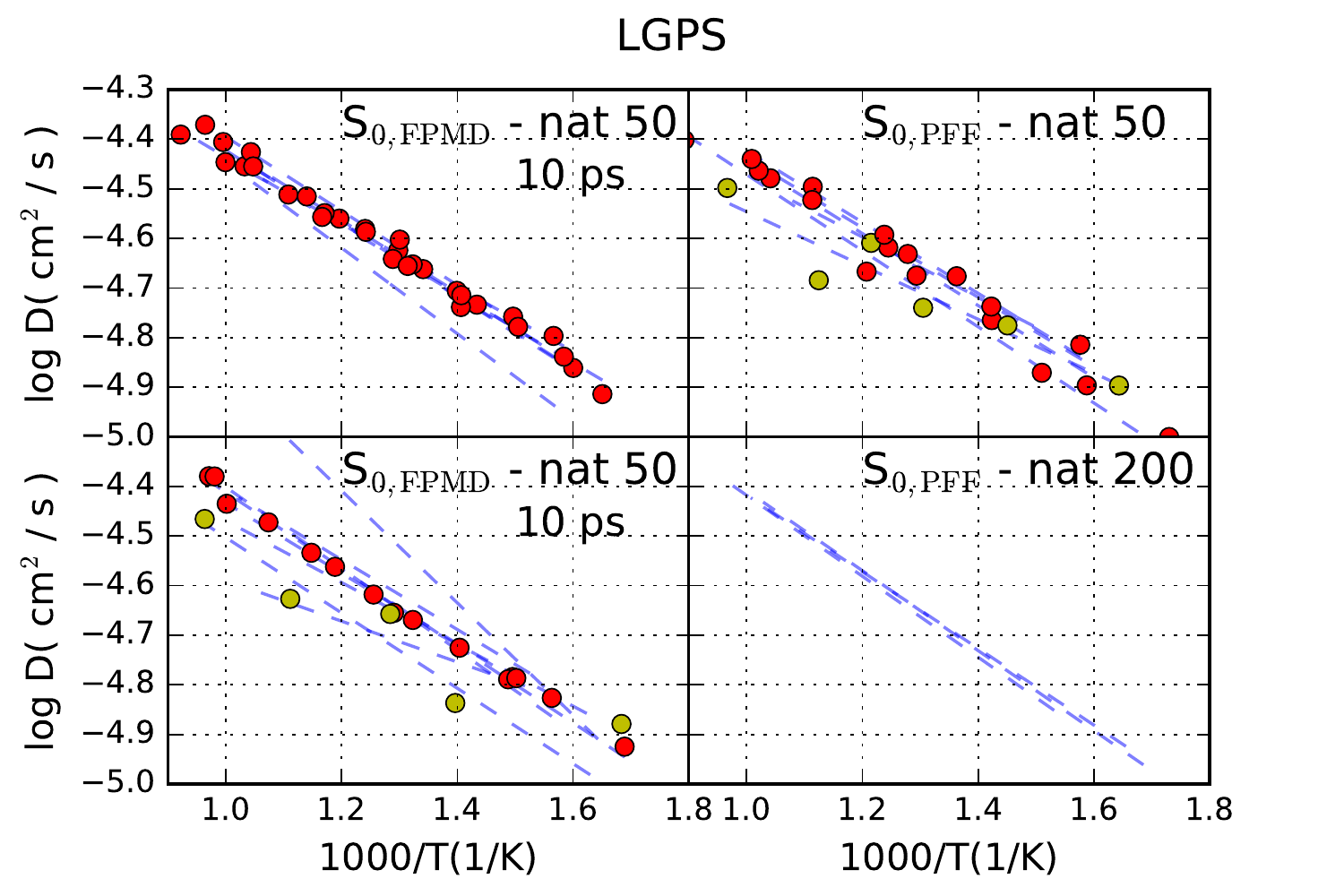}
\caption{Arrhenius plots for Li diffusion in LGPS. On the left, S0 is sampled from short FPMD
trajectories and a cell of 50 atoms, and on the right from long PFF-MD trajectories and a cell of
50 atoms (top) and 200 atoms (bottom). In all plots, the blue lines are fitted from diffusion data
of the first training loop; red points are evaluated during the second loop; yellow points belong to
the second training loop and are considered outliers. For the simulation with 200 atoms only the
first loop is reported.}
\label{fig::arr_lgps}
\end{figure}
\section{Results}
We found it necessary to apply a cleaning procedure to the training sets to help the algorithm converge to a physical model. In particular we considered the following cleaning steps:
\begin{itemize}
    \item \emph{Minimum ion-ion distance. } We rejected MD trajectories where the minimum ion-ion distance became smaller than a cutoff chosen at $0.6$ \AA. In such a case, the training step was performed again with an increased batch size for DeePMD to converge to a different, and ideally more physical, model.
    \item \emph{Energy and force cleaning. }
     We excluded a configuration as an outlier when its total energy was outside of the interval $(m-i*f,m+i*f)$, where $m$ is the median energy across all configurations of the current training set, $i$ the corresponding interquartile range, and the factor $f=2$ was manually optimized. For the forces we followed a similar procedure with $f=10$.
    \item \emph{Loss condition. } If during the training procedure the loss function $\mathcal{L}_{\omega}$ on the training set did not decrease fast enough, training was restarted from scratch with an increased batch size. This way the stochastic part of the gradient descent algorithm was reduced, permitting to reach a local minimum of the parameter space.
    \item \emph{Optional: Maximum bond variation. } If a bond is stretched or compressed more than $20 \%$ with respect to the original value, the configuration was rejected. We will explicitly state when the criterion of maximum bond variation was active in the following examples.
\end{itemize}
The evaluation of the diffusion coefficient was performed using the LAMMPS \cite{Plimpton1995} implementation of the deepMD model \cite{deepmd} under constant particle number, volume and energy conditions (NVE). The DeePMD model provided excellent energy conservation in all systems up to 2 ns of dynamics, which permitted to evaluate the Arrhenius plot for diffusion in the high temperature regime, i.e. for $T$ larger than $600$\,K, a standard choice in the field of SSE simulations. The cell used for the NVE simulations contained 200, 192 and 480 atoms for LGPS, LLZO, and NASICON, respectively. From a structural point of view LGPS, LLZO and NASICON are composed by a rigid matrix of stable units, like PS$_4$, GeS$_4$, SiO$_4$, PO$_4$ tetrahedra or ZrO$_6$ octahedra through which ionized lithium can diffuse. The models provided by our training strategy, often already during the first training loop, preserved these bonded units. Representative mean squared displacements (MSD), resolved by species, are reported in Fig. \ref{fig::all_msd}. The superionic behavior is clearly observed in all materials, with the MSD
of Lithium/Sodium growing linearly over time, and with a bounded MSD for all elements belonging to the rigid matrix of the material, which are just vibrating
around their equilibrium positions. A detailed discussion of the Arrhenius plots for each SSE follows.\\

\textbf{LGPS. } In terms of ionic conductivity, LGPS is one of the most performing SSEs, with values comparable to the ones of organic liquid electrolytes \cite{kamaya_lithium_2011}. This property is believed to be linked to the high polarizability of the sulphur atoms. The value of the diffusion coefficient for different starting training sets $S_0$ is reported in Fig. \ref{fig::arr_lgps}. On the left we report two different independent runs whose starting training sets were taken from short FPMD dynamics $10$ ps long in a supercell of $50$ atoms. 
On the right we report the diffusion coefficients resulting from PFF-MD initial configurations and using a simulation cell of $50$ (top) and $200$ (bottom) atoms, respectively.\\
Considering the runs started from FPMD configurations, a poor performance of the model is observed during the first training loop, reported as blue fitted lines, but both runs converge to very similar values when the diffusion coefficients from the second loop are considered. Only one model produced some outliers in the second loop,which we report as yellow points. For the training loops initiated by the PFF-MD initial data, we observed, in the case of the $50$ atom cell, a similar degree of stochasticity, whereas for the $200$ atom cell convergence was achieved much faster, even during the first training loop.\\
From linear fits of the Arrhenius plots, we estimate an activation energy of $0.16$ eV, which is slightly lower than experimental values ($0.22$ eV)  \cite{C3EE41728J} and in line with previous computational studies ($0.17$ and $0.21$ eV)\cite{,marcolongo_ionic_2017,ong_phase_2013}. We note that for this system the distribution of Ge and P across the simulation cell can have an influence \cite{2016_Bhattacharya_Wolverton_Carbon_Sci_Tech}. 

\begin{figure}
\includegraphics[width=\hsize]{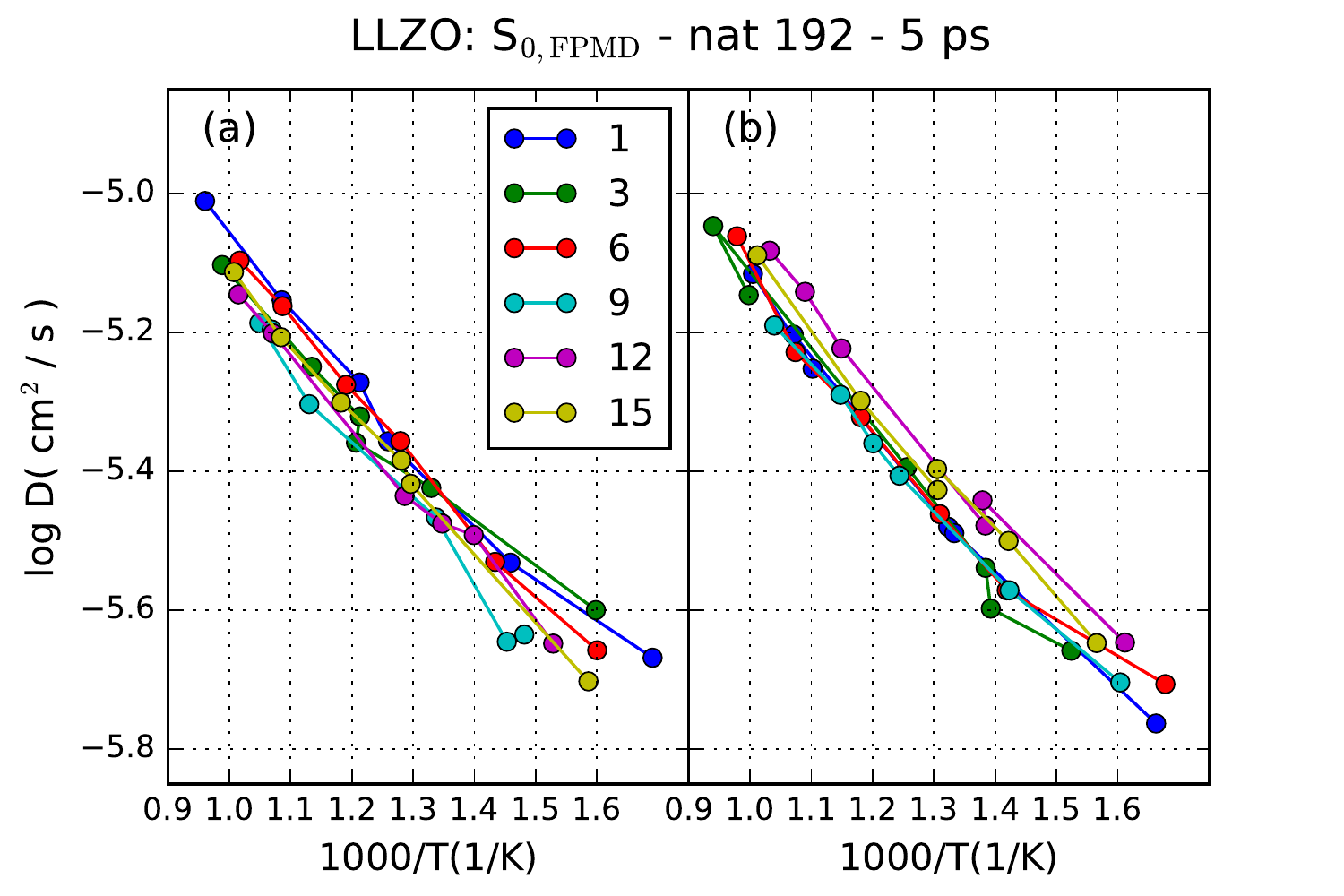}
\caption{Arrhenius plots for Li diffusion in LLZO, with $S_0$ sampled from a short FPMD trajectory.
Plots (a) and (b) report diffusion coefficients evaluated during the second and third training loop,
respectively. The color coding refers to the step number during the temperature ramp, counting
as first the training step at 300K. }
\label{fig::arr_llzo_1}
\end{figure}

\begin{figure}
\includegraphics[width=\hsize]{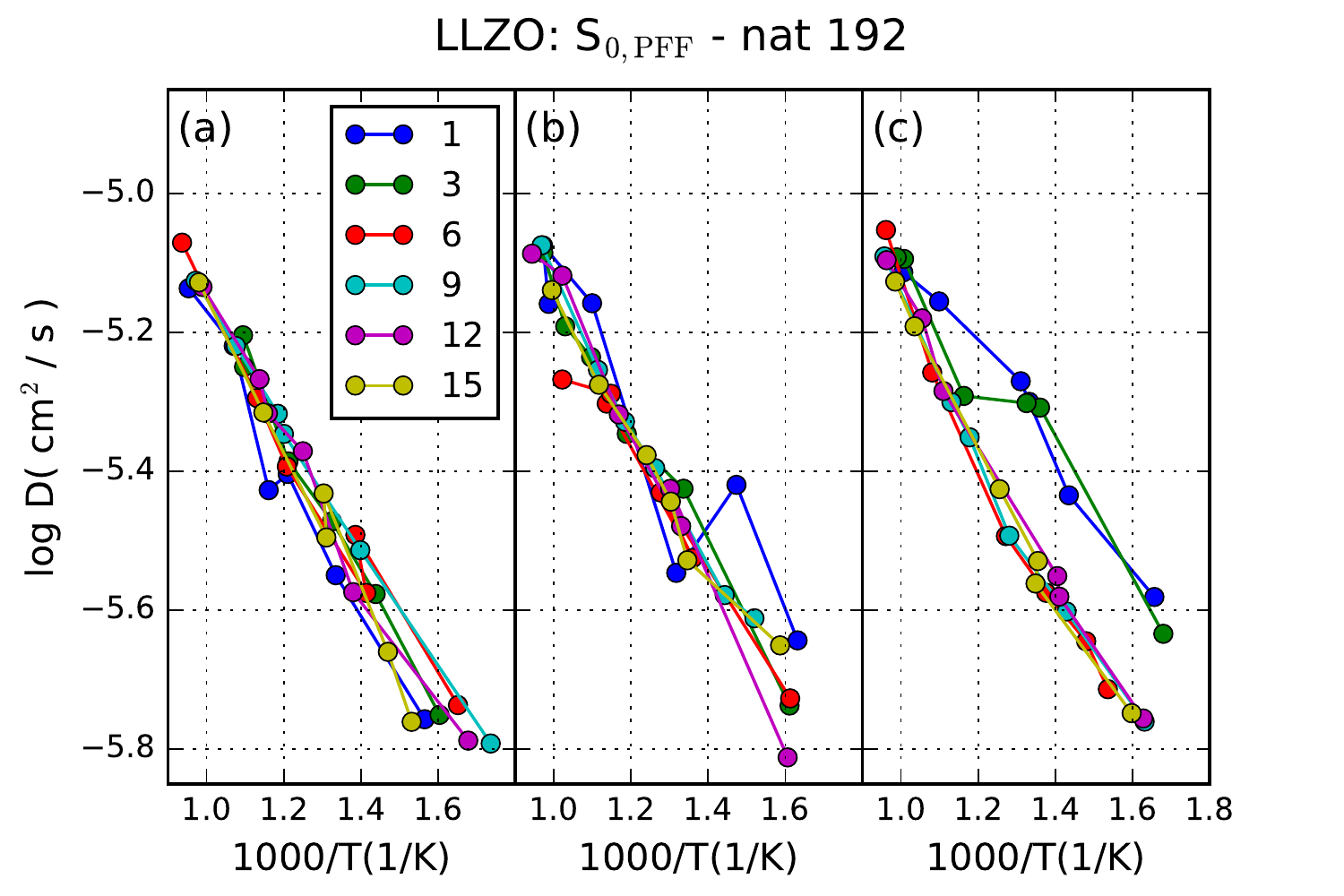}
\caption{Arrhenius plots for Li diffusion in LLZO, with $S_0$ sampled from initial PFF-MD simulations.
Only the second loops are reported and the same conventions as Fig. 4 are followed. (a,b) Training
workflows using different LLZO volumes, cf. text. (c) Diffusion coefficients computed during a
work
ow where the "maximum bond variation" check was active.}
\label{fig::arr_llzo_2}
\end{figure}

\textbf{LLZO. } LLZO can be considered as a reference material for computational studies of oxide Li-ion conducting SSEs. We included this SSE in the set of materials tested because of the chemical difference between sulphides and oxides. \\
The smallest possible simulation cell contains 192 atoms, which was chosen for all simulations. In the case of a $S_0$ set sampled from a short, $5$ps long, FPMD trajectory, convergence was obtained during the second training loop. In Fig. \ref{fig::arr_llzo_1}, we report the Arrhenius plots evaluated during the second and third loops of the training workflow. Overall the spread of the computed diffusion coefficients is around $\sim 0.1$ in Log-scale, which corresponds to a factor of $\sim 1.25$. Such small shift is not important for applications and could be related to a different modelling of the vibrational frequencies. The activation energies are very stable with values ranging from $0.2$ to $0.22$ eV. Previous computational studies of LLZO report values between $0.18$ and $0.27$ eV \cite{PhysRevMaterials.3.035403,kozinsky_effects_2016,DAI2019333}. Our results are in very good agreement with the experimental activation energy of cubic LLZO. High temperature experimental measurements give values of $0.21$ and $0.20$ eV \cite{WANG2015612, LLZO_exp}. Note that the reported value of the activation energy reported in \cite{LLZO_exp} is lower because it was fitted on conductivity rather than on diffusion data. Once converting conductivity to diffusivity via the Nernst-Einstein equation \cite{marcolongo_ionic_2017} one recovers compatibility between experimental results and the ones computed in this work. \\
In the case of a $S_0$ set sampled from long PFF-MD simulations, we obtained similar values of the diffusion coefficient and activation energies. For LLZO, we tested two different volumes of the cell chosen to produce $S_0$. In the first case, we equilibrated with NPT simulations, similar to the procedure followed for the other materials. In the second case, we used the experimental value as reported in the cif-file. The results are quite insensitive with respect to this choice, as can be seen in the first and second plot of Fig. \ref{fig::arr_llzo_2} .
Interestingly, when switching on the ``maximum bond variation'' criterion, we observed a wrong activation energy for the first training steps of the second loop, as can be seen in the third plot of Fig. \ref{fig::arr_llzo_2}. Nevertheless, after reaching convergence, all diffusion coefficients were compatible. This result suggests that inserting unfavourable configurations in the training set can provide a speed-up of convergence because the potential energy model can be better interpolated. During the last training steps, when the trajctories computed were more physical, configurations were not rejected any more by the cleaning procedure. Further, we observe that during the first training steps it is much harder to recognize an Arrhenius-like behavior, i.e. a line in semilog scale with a definite activation energy. As soon as the models become smoother thanks to an enlarged of training set, the Arrhenius behavior is recovered. 

\

\begin{figure}
\includegraphics[width=\hsize]{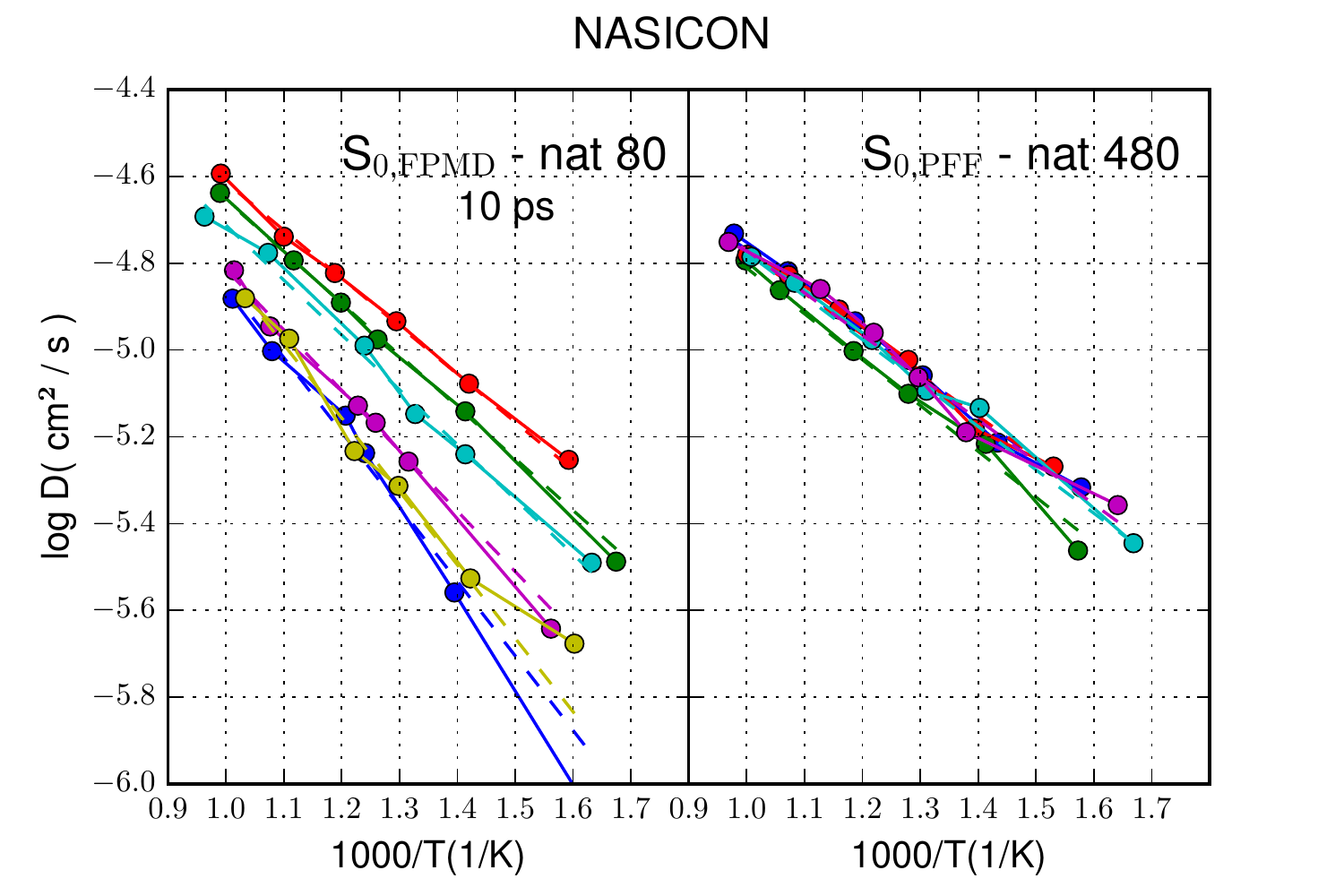}
\caption{Arrhenius plots for Na diffusion in NASICON with two different choices of $S_0$. For the
initial training set from a FPMD trajectory with an 80 atoms cell, we report data evaluated during
the third training loop at different training steps. No convergence is observed. When $S_0$ is sampled
from PFF-MD simulations and a larger cell (480 atoms), convergence is achieved already during
the first training loop.
}
\label{fig::all_nasicon}
\end{figure}

\textbf{NASICON. } We tested NASICON as a representative of Na-ion conducting SSE materials. Among the NASICON family Na$_{1+x}$Zr$_2$Si$_x$P$_{3-x}$O$_{12}$, the member with $x=2$ experimentally shows the highest value of ionic conductivity \cite{NASICON_2} and is therefore used for this study. We produced the initial training set $S_0$ from a $10$ps long trajectory via FPMD with a small simulation cell of 80 atoms. Using this choice of $S_0$, almost all DeePMD models trained during the first loop failed to produce meaningful NVE dynamics for the evaluation of the diffusion coefficient due to non-physical trajectories. Nevertheless, the thermostat active in NVT MD during the training steps produced physical trajectories so that the training workflow could proceed. Consequently, we obtained Arrhenius plots of the diffusion coefficient during the second and third training loop. However, also after introducing the different cleaning procedures, we could not observe a clear convergence of the workflow from the Arrhenius plots even after the third loop, when more than 3000 configurations had been added to the training set. This is exemplified in the left of Fig. \ref{fig::all_nasicon}.  The fitted activation energies spanned a range from $0.2$ to $0.4$ eV, which represents a large error resulting in large differences when extrapolating the diffusion coefficient to ambient temperature. In contrast, when the training loop is started with an initial training set from PFF-MD configurations and a larger simulation cell (480 atoms), convergence is obtained even during the first training loop. The corresponding activation energy evaluated from the right plot in Fig. \ref{fig::all_nasicon} was around $0.2$ eV, in line with high temperature experiments providing activation energies between $0.18$ and $0.20$ eV \cite{NASICON_2}. \\
We note that the number of force components in $S_0$ was equal to $\sim 4*10^5$ for PFF-MD and $\sim 3*10^4$ for FPMD (see Appendix A). During the third training loop started from
a short FPMD trajectory and reported in Fig. \ref{fig::all_nasicon}, the number of force components in the
training set more than doubled the force components initially present in the training set
$S_0$ sampled from PFF-FPMD. Nevertheless, the diffusion coefficients struggled to converge. This example showcases the importance of starting the training loop with enough data which sample a relevant portion of the phase space in order to obtain generalizable models. \\ 

\section{Conclusions}
We conclude that the DeePMD model is a good candidate to simulate transport in SSEs, able to reproduce the superionic behavior and to provide reasonable activation energies. Nevertheless, care must be taken when choosing the training set. The proposed training protocol based on an augmenting training set via intermediate molecular dynamics simulations is able to assess the quality of the approximate trajectories and of the computed physical properties. We found it important to perform a cleaning of the training set to remove outliers which can impede the training of a physically meaningful model. The starting set of configurations $S_0$ can be extracted either from short FPMD simulations or using an approximate classical PFF potential. We found that the latter choice led to a faster convergence of the training workflow, because larger simulation cells could be used.\\
In this work we trained an independent model for each material. Building a universal potential in future work will require the weights, i.e. the parameters, of the neural network to be shared and trained on a larger set of materials. Our results show that the DeePMD model is promising for developing such a universal potential, even if care must be taken when selecting the configurations used for training.

\begin{acknowledgements}
This research was supported by the NCCR MARVEL, funded by the Swiss National Science Foundation. This work was supported by a grant from the Swiss National Supercomputing Centre (CSCS) under project ID mr28. A.M would like to thank Linfeng Zhang for help in the usage of the DeePMD code.
\end{acknowledgements}

\appendix

\section{Simulation parameters}
The short FPMD simulations were performed at the PBE level \cite{perdew_generalized_1996} of DFT including van-der-Waals correction within the Grimme-D2 parametrization \cite{grimme_semiempirical_2006} as implemented in the CPMD code \cite{CPMD}. The corresponding structures were taken from ICSD \cite{belsky_new_2002} file numbers 188886 , 422259, and 473 for LGPS, LLZO, and NASICON, respectively.\\
The parameters of the supercells used for the NVE calculations were taken from Ref. \cite{C3EE41728J} for LGPS, from ICSD entry 422259 for LLZO, and from ICSD entry 62383 for NASICON. \\ 
The total number of configurations in the starting training set $S_0$ was equal to 240, 60, and 120 for LGPS, LLZO, and NASICON, respectively, selected from the FPMD simulations. Increasing the number of configurations, i.e. increasing the sampling frequency, did not improve the convergence of the workflow. We conclude that using uncorrelated snapshots provides the best performance. The starting training sets $S_0$ selected from PFF-MD contained a total number of $300$ uncorrelated snapshots. \\
For the training workflows started from FPMD initial data, 100 configurations were added at every training step, sampling every picosecond the NVT simulations. For the training workflows started from PFF-MD initial data, only 40 configurations were added at every training step, sampled with the same frequency, because of the larger simulation cells used.\\
All parameters of the DeePMD model were fixed in this work to the suggested default values
and the virial was introduced in the loss function. A training batch size of 1 was generally used, which was increased if a training step had to be repeated.

\bibliography{biblio.bib}

\begin{thebibliography}{43}%
\makeatletter
\providecommand \@ifxundefined [1]{%
 \@ifx{#1\undefined}
}%
\providecommand \@ifnum [1]{%
 \ifnum #1\expandafter \@firstoftwo
 \else \expandafter \@secondoftwo
 \fi
}%
\providecommand \@ifx [1]{%
 \ifx #1\expandafter \@firstoftwo
 \else \expandafter \@secondoftwo
 \fi
}%
\providecommand \natexlab [1]{#1}%
\providecommand \enquote  [1]{``#1''}%
\providecommand \bibnamefont  [1]{#1}%
\providecommand \bibfnamefont [1]{#1}%
\providecommand \citenamefont [1]{#1}%
\providecommand \href@noop [0]{\@secondoftwo}%
\providecommand \href [0]{\begingroup \@sanitize@url \@href}%
\providecommand \@href[1]{\@@startlink{#1}\@@href}%
\providecommand \@@href[1]{\endgroup#1\@@endlink}%
\providecommand \@sanitize@url [0]{\catcode `\\12\catcode `\$12\catcode
  `\&12\catcode `\#12\catcode `\^12\catcode `\_12\catcode `\%12\relax}%
\providecommand \@@startlink[1]{}%
\providecommand \@@endlink[0]{}%
\providecommand \url  [0]{\begingroup\@sanitize@url \@url }%
\providecommand \@url [1]{\endgroup\@href {#1}{\urlprefix }}%
\providecommand \urlprefix  [0]{URL }%
\providecommand \Eprint [0]{\href }%
\providecommand \doibase [0]{http://dx.doi.org/}%
\providecommand \selectlanguage [0]{\@gobble}%
\providecommand \bibinfo  [0]{\@secondoftwo}%
\providecommand \bibfield  [0]{\@secondoftwo}%
\providecommand \translation [1]{[#1]}%
\providecommand \BibitemOpen [0]{}%
\providecommand \bibitemStop [0]{}%
\providecommand \bibitemNoStop [0]{.\EOS\space}%
\providecommand \EOS [0]{\spacefactor3000\relax}%
\providecommand \BibitemShut  [1]{\csname bibitem#1\endcsname}%
\let\auto@bib@innerbib\@empty
\bibitem [{\citenamefont {Zhang}\ \emph {et~al.}(2018)\citenamefont {Zhang},
  \citenamefont {Han}, \citenamefont {Wang}, \citenamefont {Car},\ and\
  \citenamefont {E}}]{PhysRevLett.120.143001}%
  \BibitemOpen
  \bibfield  {author} {\bibinfo {author} {\bibfnamefont {L.}~\bibnamefont
  {Zhang}}, \bibinfo {author} {\bibfnamefont {J.}~\bibnamefont {Han}}, \bibinfo
  {author} {\bibfnamefont {H.}~\bibnamefont {Wang}}, \bibinfo {author}
  {\bibfnamefont {R.}~\bibnamefont {Car}}, \ and\ \bibinfo {author}
  {\bibfnamefont {W.}~\bibnamefont {E}},\ }\href {\doibase
  10.1103/PhysRevLett.120.143001} {\bibfield  {journal} {\bibinfo  {journal}
  {Phys. Rev. Lett.}\ }\textbf {\bibinfo {volume} {120}},\ \bibinfo {pages}
  {143001} (\bibinfo {year} {2018})}\BibitemShut {NoStop}%
\bibitem [{\citenamefont {Kim}\ \emph {et~al.}(2015)\citenamefont {Kim},
  \citenamefont {Son}, \citenamefont {Mukherjee}, \citenamefont {Schuppert},
  \citenamefont {Bates}, \citenamefont {Kwon}, \citenamefont {Choi},
  \citenamefont {Chung},\ and\ \citenamefont {Park}}]{kim_review_2015}%
  \BibitemOpen
  \bibfield  {author} {\bibinfo {author} {\bibfnamefont {J.~G.}\ \bibnamefont
  {Kim}}, \bibinfo {author} {\bibfnamefont {B.}~\bibnamefont {Son}}, \bibinfo
  {author} {\bibfnamefont {S.}~\bibnamefont {Mukherjee}}, \bibinfo {author}
  {\bibfnamefont {N.}~\bibnamefont {Schuppert}}, \bibinfo {author}
  {\bibfnamefont {A.}~\bibnamefont {Bates}}, \bibinfo {author} {\bibfnamefont
  {O.}~\bibnamefont {Kwon}}, \bibinfo {author} {\bibfnamefont {M.~J.}\
  \bibnamefont {Choi}}, \bibinfo {author} {\bibfnamefont {H.~Y.}\ \bibnamefont
  {Chung}}, \ and\ \bibinfo {author} {\bibfnamefont {S.}~\bibnamefont {Park}},\
  }\href {\doibase 10.1016/j.jpowsour.2015.02.054} {\bibfield  {journal}
  {\bibinfo  {journal} {Journal of Power Sources}\ }\textbf {\bibinfo {volume}
  {282}},\ \bibinfo {pages} {299} (\bibinfo {year} {2015})}\BibitemShut
  {NoStop}%
\bibitem [{\citenamefont {Weppner}(2003)}]{weppner_engineering_2003}%
  \BibitemOpen
  \bibfield  {author} {\bibinfo {author} {\bibfnamefont {W.}~\bibnamefont
  {Weppner}},\ }\href {\doibase 10.1007/BF02376599} {\bibfield  {journal}
  {\bibinfo  {journal} {Ionics}\ }\textbf {\bibinfo {volume} {9}},\ \bibinfo
  {pages} {444} (\bibinfo {year} {2003})}\BibitemShut {NoStop}%
\bibitem [{\citenamefont {Tikekar}\ \emph {et~al.}(2016)\citenamefont
  {Tikekar}, \citenamefont {Choudhury}, \citenamefont {Tu},\ and\ \citenamefont
  {Archer}}]{tikekar_design_2016}%
  \BibitemOpen
  \bibfield  {author} {\bibinfo {author} {\bibfnamefont {M.~D.}\ \bibnamefont
  {Tikekar}}, \bibinfo {author} {\bibfnamefont {S.}~\bibnamefont {Choudhury}},
  \bibinfo {author} {\bibfnamefont {Z.}~\bibnamefont {Tu}}, \ and\ \bibinfo
  {author} {\bibfnamefont {L.~A.}\ \bibnamefont {Archer}},\ }\href {\doibase
  10.1038/nenergy.2016.114} {\bibfield  {journal} {\bibinfo  {journal} {Nature
  Energy}\ }\textbf {\bibinfo {volume} {1}},\ \bibinfo {pages} {16114}
  (\bibinfo {year} {2016})}\BibitemShut {NoStop}%
\bibitem [{\citenamefont {Frenkel}\ and\ \citenamefont
  {Smit}(1996)}]{Frenkel:1996:UMS:547952}%
  \BibitemOpen
  \bibinfo {editor} {\bibfnamefont {D.}~\bibnamefont {Frenkel}}\ and\ \bibinfo
  {editor} {\bibfnamefont {B.}~\bibnamefont {Smit}},\ eds.,\ \href@noop {}
  {\emph {\bibinfo {title} {Understanding Molecular Simulation: From Algorithms
  to Applications}}},\ \bibinfo {edition} {1st}\ ed.\ (\bibinfo  {publisher}
  {Academic Press, Inc.},\ \bibinfo {address} {Orlando, FL, USA},\ \bibinfo
  {year} {1996})\BibitemShut {NoStop}%
\bibitem [{\citenamefont {Muy}\ \emph {et~al.}(2019)\citenamefont {Muy},
  \citenamefont {Voss}, \citenamefont {Schlem}, \citenamefont {Koerver},
  \citenamefont {Sedlmaier}, \citenamefont {Maglia}, \citenamefont {Lamp},
  \citenamefont {Zeier},\ and\ \citenamefont
  {Shao-Horn}}]{muy_high-throughput_2019}%
  \BibitemOpen
  \bibfield  {author} {\bibinfo {author} {\bibfnamefont {S.}~\bibnamefont
  {Muy}}, \bibinfo {author} {\bibfnamefont {J.}~\bibnamefont {Voss}}, \bibinfo
  {author} {\bibfnamefont {R.}~\bibnamefont {Schlem}}, \bibinfo {author}
  {\bibfnamefont {R.}~\bibnamefont {Koerver}}, \bibinfo {author} {\bibfnamefont
  {S.~J.}\ \bibnamefont {Sedlmaier}}, \bibinfo {author} {\bibfnamefont
  {F.}~\bibnamefont {Maglia}}, \bibinfo {author} {\bibfnamefont
  {P.}~\bibnamefont {Lamp}}, \bibinfo {author} {\bibfnamefont {W.~G.}\
  \bibnamefont {Zeier}}, \ and\ \bibinfo {author} {\bibfnamefont
  {Y.}~\bibnamefont {Shao-Horn}},\ }\href {\doibase 10.1016/j.isci.2019.05.036}
  {\bibfield  {journal} {\bibinfo  {journal} {iScience}\ }\textbf {\bibinfo
  {volume} {16}},\ \bibinfo {pages} {270} (\bibinfo {year} {2019})}\BibitemShut
  {NoStop}%
\bibitem [{\citenamefont {Sendek}\ \emph {et~al.}(2017)\citenamefont {Sendek},
  \citenamefont {Yang}, \citenamefont {Cubuk}, \citenamefont {Duerloo},
  \citenamefont {Cui},\ and\ \citenamefont {Reed}}]{d.sendek_holistic_2017}%
  \BibitemOpen
  \bibfield  {author} {\bibinfo {author} {\bibfnamefont {A.~D.}\ \bibnamefont
  {Sendek}}, \bibinfo {author} {\bibfnamefont {Q.}~\bibnamefont {Yang}},
  \bibinfo {author} {\bibfnamefont {E.~D.}\ \bibnamefont {Cubuk}}, \bibinfo
  {author} {\bibfnamefont {K.-A.~N.}\ \bibnamefont {Duerloo}}, \bibinfo
  {author} {\bibfnamefont {Y.}~\bibnamefont {Cui}}, \ and\ \bibinfo {author}
  {\bibfnamefont {E.~J.}\ \bibnamefont {Reed}},\ }\href {\doibase
  10.1039/C6EE02697D} {\bibfield  {journal} {\bibinfo  {journal} {Energy \&
  Environmental Science}\ }\textbf {\bibinfo {volume} {10}},\ \bibinfo {pages}
  {306} (\bibinfo {year} {2017})}\BibitemShut {NoStop}%
\bibitem [{\citenamefont {Gao}\ \emph {et~al.}(2014)\citenamefont {Gao},
  \citenamefont {Chu}, \citenamefont {He}, \citenamefont {Zhang}, \citenamefont
  {Xiao}, \citenamefont {Li},\ and\ \citenamefont {Chen}}]{Gao2014}%
  \BibitemOpen
  \bibfield  {author} {\bibinfo {author} {\bibfnamefont {J.}~\bibnamefont
  {Gao}}, \bibinfo {author} {\bibfnamefont {G.}~\bibnamefont {Chu}}, \bibinfo
  {author} {\bibfnamefont {M.}~\bibnamefont {He}}, \bibinfo {author}
  {\bibfnamefont {S.}~\bibnamefont {Zhang}}, \bibinfo {author} {\bibfnamefont
  {R.}~\bibnamefont {Xiao}}, \bibinfo {author} {\bibfnamefont {H.}~\bibnamefont
  {Li}}, \ and\ \bibinfo {author} {\bibfnamefont {L.}~\bibnamefont {Chen}},\
  }\href {\doibase 10.1007/s11433-014-5511-4} {\bibfield  {journal} {\bibinfo
  {journal} {Science China Physics, Mechanics {\&} Astronomy}\ }\textbf
  {\bibinfo {volume} {57}},\ \bibinfo {pages} {1526} (\bibinfo {year}
  {2014})}\BibitemShut {NoStop}%
\bibitem [{\citenamefont {Wang}\ \emph {et~al.}(2015)\citenamefont {Wang},
  \citenamefont {Richards}, \citenamefont {Ong}, \citenamefont {Miara},
  \citenamefont {Kim}, \citenamefont {Mo},\ and\ \citenamefont
  {Ceder}}]{wang_design_2015}%
  \BibitemOpen
  \bibfield  {author} {\bibinfo {author} {\bibfnamefont {Y.}~\bibnamefont
  {Wang}}, \bibinfo {author} {\bibfnamefont {W.~D.}\ \bibnamefont {Richards}},
  \bibinfo {author} {\bibfnamefont {S.~P.}\ \bibnamefont {Ong}}, \bibinfo
  {author} {\bibfnamefont {L.~J.}\ \bibnamefont {Miara}}, \bibinfo {author}
  {\bibfnamefont {J.~C.}\ \bibnamefont {Kim}}, \bibinfo {author} {\bibfnamefont
  {Y.}~\bibnamefont {Mo}}, \ and\ \bibinfo {author} {\bibfnamefont
  {G.}~\bibnamefont {Ceder}},\ }\href {\doibase 10.1038/nmat4369} {\bibfield
  {journal} {\bibinfo  {journal} {Nature Materials}\ }\textbf {\bibinfo
  {volume} {14}},\ \bibinfo {pages} {1026} (\bibinfo {year}
  {2015})}\BibitemShut {NoStop}%
\bibitem [{\citenamefont {Kahle}\ \emph {et~al.}(2018)\citenamefont {Kahle},
  \citenamefont {Marcolongo},\ and\ \citenamefont
  {Marzari}}]{kahle_modeling_2018}%
  \BibitemOpen
  \bibfield  {author} {\bibinfo {author} {\bibfnamefont {L.}~\bibnamefont
  {Kahle}}, \bibinfo {author} {\bibfnamefont {A.}~\bibnamefont {Marcolongo}}, \
  and\ \bibinfo {author} {\bibfnamefont {N.}~\bibnamefont {Marzari}},\ }\href
  {\doibase 10.1103/PhysRevMaterials.2.065405} {\bibfield  {journal} {\bibinfo
  {journal} {Physical Review Materials}\ }\textbf {\bibinfo {volume} {2}},\
  \bibinfo {pages} {065405} (\bibinfo {year} {2018})}\BibitemShut {NoStop}%
\bibitem [{\citenamefont {Goodfellow}\ \emph {et~al.}(2016)\citenamefont
  {Goodfellow}, \citenamefont {Bengio},\ and\ \citenamefont
  {Courville}}]{Goodfellow:2016:DL:3086952}%
  \BibitemOpen
  \bibfield  {author} {\bibinfo {author} {\bibfnamefont {I.}~\bibnamefont
  {Goodfellow}}, \bibinfo {author} {\bibfnamefont {Y.}~\bibnamefont {Bengio}},
  \ and\ \bibinfo {author} {\bibfnamefont {A.}~\bibnamefont {Courville}},\
  }\href@noop {} {\emph {\bibinfo {title} {Deep Learning}}}\ (\bibinfo
  {publisher} {The MIT Press},\ \bibinfo {year} {2016})\BibitemShut {NoStop}%
\bibitem [{\citenamefont {Behler}\ and\ \citenamefont
  {Parrinello}(2007)}]{PhysRevLett.98.146401}%
  \BibitemOpen
  \bibfield  {author} {\bibinfo {author} {\bibfnamefont {J.}~\bibnamefont
  {Behler}}\ and\ \bibinfo {author} {\bibfnamefont {M.}~\bibnamefont
  {Parrinello}},\ }\href {\doibase 10.1103/PhysRevLett.98.146401} {\bibfield
  {journal} {\bibinfo  {journal} {Phys. Rev. Lett.}\ }\textbf {\bibinfo
  {volume} {98}},\ \bibinfo {pages} {146401} (\bibinfo {year}
  {2007})}\BibitemShut {NoStop}%
\bibitem [{\citenamefont {Bart\'ok}\ \emph {et~al.}(2010)\citenamefont
  {Bart\'ok}, \citenamefont {Payne}, \citenamefont {Kondor},\ and\
  \citenamefont {Cs\'anyi}}]{PhysRevLett.104.136403}%
  \BibitemOpen
  \bibfield  {author} {\bibinfo {author} {\bibfnamefont {A.~P.}\ \bibnamefont
  {Bart\'ok}}, \bibinfo {author} {\bibfnamefont {M.~C.}\ \bibnamefont {Payne}},
  \bibinfo {author} {\bibfnamefont {R.}~\bibnamefont {Kondor}}, \ and\ \bibinfo
  {author} {\bibfnamefont {G.}~\bibnamefont {Cs\'anyi}},\ }\href {\doibase
  10.1103/PhysRevLett.104.136403} {\bibfield  {journal} {\bibinfo  {journal}
  {Phys. Rev. Lett.}\ }\textbf {\bibinfo {volume} {104}},\ \bibinfo {pages}
  {136403} (\bibinfo {year} {2010})}\BibitemShut {NoStop}%
\bibitem [{\citenamefont {Li}\ \emph {et~al.}(2015)\citenamefont {Li},
  \citenamefont {Kermode},\ and\ \citenamefont
  {De~Vita}}]{PhysRevLett.114.096405}%
  \BibitemOpen
  \bibfield  {author} {\bibinfo {author} {\bibfnamefont {Z.}~\bibnamefont
  {Li}}, \bibinfo {author} {\bibfnamefont {J.~R.}\ \bibnamefont {Kermode}}, \
  and\ \bibinfo {author} {\bibfnamefont {A.}~\bibnamefont {De~Vita}},\ }\href
  {\doibase 10.1103/PhysRevLett.114.096405} {\bibfield  {journal} {\bibinfo
  {journal} {Phys. Rev. Lett.}\ }\textbf {\bibinfo {volume} {114}},\ \bibinfo
  {pages} {096405} (\bibinfo {year} {2015})}\BibitemShut {NoStop}%
\bibitem [{\citenamefont {Botu}\ and\ \citenamefont
  {Ramprasad}(2015)}]{botu_094306}%
  \BibitemOpen
  \bibfield  {author} {\bibinfo {author} {\bibfnamefont {V.}~\bibnamefont
  {Botu}}\ and\ \bibinfo {author} {\bibfnamefont {R.}~\bibnamefont
  {Ramprasad}},\ }\href {\doibase 10.1103/PhysRevB.92.094306} {\bibfield
  {journal} {\bibinfo  {journal} {Phys. Rev. B}\ }\textbf {\bibinfo {volume}
  {92}},\ \bibinfo {pages} {094306} (\bibinfo {year} {2015})}\BibitemShut
  {NoStop}%
\bibitem [{\citenamefont {Hellström}\ and\ \citenamefont
  {Behler}(2016)}]{doi:10.1021/acs.jpclett.6b01448}%
  \BibitemOpen
  \bibfield  {author} {\bibinfo {author} {\bibfnamefont {M.}~\bibnamefont
  {Hellström}}\ and\ \bibinfo {author} {\bibfnamefont {J.}~\bibnamefont
  {Behler}},\ }\href {\doibase 10.1021/acs.jpclett.6b01448} {\bibfield
  {journal} {\bibinfo  {journal} {The Journal of Physical Chemistry Letters}\
  }\textbf {\bibinfo {volume} {7}},\ \bibinfo {pages} {3302} (\bibinfo {year}
  {2016})},\ \bibinfo {note} {pMID: 27504986},\ \Eprint
  {http://arxiv.org/abs/https://doi.org/10.1021/acs.jpclett.6b01448}
  {https://doi.org/10.1021/acs.jpclett.6b01448} \BibitemShut {NoStop}%
\bibitem [{\citenamefont {Smith}\ \emph {et~al.}(2017)\citenamefont {Smith},
  \citenamefont {Isayev},\ and\ \citenamefont {Roitberg}}]{C6SC05720A}%
  \BibitemOpen
  \bibfield  {author} {\bibinfo {author} {\bibfnamefont {J.~S.}\ \bibnamefont
  {Smith}}, \bibinfo {author} {\bibfnamefont {O.}~\bibnamefont {Isayev}}, \
  and\ \bibinfo {author} {\bibfnamefont {A.~E.}\ \bibnamefont {Roitberg}},\
  }\href {\doibase 10.1039/C6SC05720A} {\bibfield  {journal} {\bibinfo
  {journal} {Chem. Sci.}\ }\textbf {\bibinfo {volume} {8}},\ \bibinfo {pages}
  {3192} (\bibinfo {year} {2017})}\BibitemShut {NoStop}%
\bibitem [{\citenamefont {{Wang}}\ \emph {et~al.}(2018)\citenamefont {{Wang}},
  \citenamefont {{Zhang}}, \citenamefont {{Han}},\ and\ \citenamefont
  {{E}}}]{2018CoPhC.228..178W}%
  \BibitemOpen
  \bibfield  {author} {\bibinfo {author} {\bibfnamefont {H.}~\bibnamefont
  {{Wang}}}, \bibinfo {author} {\bibfnamefont {L.}~\bibnamefont {{Zhang}}},
  \bibinfo {author} {\bibfnamefont {J.}~\bibnamefont {{Han}}}, \ and\ \bibinfo
  {author} {\bibfnamefont {W.}~\bibnamefont {{E}}},\ }\href {\doibase
  10.1016/j.cpc.2018.03.016} {\bibfield  {journal} {\bibinfo  {journal}
  {Computer Physics Communications}\ }\textbf {\bibinfo {volume} {228}},\
  \bibinfo {pages} {178} (\bibinfo {year} {2018})},\ \Eprint
  {http://arxiv.org/abs/1712.03641} {arXiv:1712.03641 [physics.comp-ph]}
  \BibitemShut {NoStop}%
\bibitem [{\citenamefont {Zhang}\ \emph {et~al.}(2019)\citenamefont {Zhang},
  \citenamefont {Lin}, \citenamefont {Wang}, \citenamefont {Car},\ and\
  \citenamefont {E}}]{PhysRevMaterials.3.023804}%
  \BibitemOpen
  \bibfield  {author} {\bibinfo {author} {\bibfnamefont {L.}~\bibnamefont
  {Zhang}}, \bibinfo {author} {\bibfnamefont {D.-Y.}\ \bibnamefont {Lin}},
  \bibinfo {author} {\bibfnamefont {H.}~\bibnamefont {Wang}}, \bibinfo {author}
  {\bibfnamefont {R.}~\bibnamefont {Car}}, \ and\ \bibinfo {author}
  {\bibfnamefont {W.}~\bibnamefont {E}},\ }\href {\doibase
  10.1103/PhysRevMaterials.3.023804} {\bibfield  {journal} {\bibinfo  {journal}
  {Phys. Rev. Materials}\ }\textbf {\bibinfo {volume} {3}},\ \bibinfo {pages}
  {023804} (\bibinfo {year} {2019})}\BibitemShut {NoStop}%
\bibitem [{\citenamefont {Kamaya}\ \emph {et~al.}(2011)\citenamefont {Kamaya},
  \citenamefont {Homma}, \citenamefont {Yamakawa}, \citenamefont {Hirayama},
  \citenamefont {Kanno}, \citenamefont {Yonemura}, \citenamefont {Kamiyama},
  \citenamefont {Kato}, \citenamefont {Hama}, \citenamefont {Kawamoto},\ and\
  \citenamefont {Mitsui}}]{kamaya_lithium_2011}%
  \BibitemOpen
  \bibfield  {author} {\bibinfo {author} {\bibfnamefont {N.}~\bibnamefont
  {Kamaya}}, \bibinfo {author} {\bibfnamefont {K.}~\bibnamefont {Homma}},
  \bibinfo {author} {\bibfnamefont {Y.}~\bibnamefont {Yamakawa}}, \bibinfo
  {author} {\bibfnamefont {M.}~\bibnamefont {Hirayama}}, \bibinfo {author}
  {\bibfnamefont {R.}~\bibnamefont {Kanno}}, \bibinfo {author} {\bibfnamefont
  {M.}~\bibnamefont {Yonemura}}, \bibinfo {author} {\bibfnamefont
  {T.}~\bibnamefont {Kamiyama}}, \bibinfo {author} {\bibfnamefont
  {Y.}~\bibnamefont {Kato}}, \bibinfo {author} {\bibfnamefont {S.}~\bibnamefont
  {Hama}}, \bibinfo {author} {\bibfnamefont {K.}~\bibnamefont {Kawamoto}}, \
  and\ \bibinfo {author} {\bibfnamefont {A.}~\bibnamefont {Mitsui}},\ }\href
  {\doibase 10.1038/nmat3066} {\bibfield  {journal} {\bibinfo  {journal}
  {Nature Materials}\ }\textbf {\bibinfo {volume} {10}},\ \bibinfo {pages}
  {682} (\bibinfo {year} {2011})}\BibitemShut {NoStop}%
\bibitem [{\citenamefont {Mo}\ \emph {et~al.}(2012)\citenamefont {Mo},
  \citenamefont {Ong},\ and\ \citenamefont {Ceder}}]{mo_first_2012}%
  \BibitemOpen
  \bibfield  {author} {\bibinfo {author} {\bibfnamefont {Y.}~\bibnamefont
  {Mo}}, \bibinfo {author} {\bibfnamefont {S.~P.}\ \bibnamefont {Ong}}, \ and\
  \bibinfo {author} {\bibfnamefont {G.}~\bibnamefont {Ceder}},\ }\href
  {\doibase 10.1021/cm203303y} {\bibfield  {journal} {\bibinfo  {journal}
  {Chemistry of Materials}\ }\textbf {\bibinfo {volume} {24}},\ \bibinfo
  {pages} {15} (\bibinfo {year} {2012})}\BibitemShut {NoStop}%
\bibitem [{\citenamefont {Marcolongo}\ and\ \citenamefont
  {Marzari}(2017)}]{marcolongo_ionic_2017}%
  \BibitemOpen
  \bibfield  {author} {\bibinfo {author} {\bibfnamefont {A.}~\bibnamefont
  {Marcolongo}}\ and\ \bibinfo {author} {\bibfnamefont {N.}~\bibnamefont
  {Marzari}},\ }\href {\doibase 10.1103/PhysRevMaterials.1.025402} {\bibfield
  {journal} {\bibinfo  {journal} {Physical Review Materials}\ }\textbf
  {\bibinfo {volume} {1}},\ \bibinfo {pages} {025402} (\bibinfo {year}
  {2017})}\BibitemShut {NoStop}%
\bibitem [{\citenamefont {Murugan}\ \emph {et~al.}(2007)\citenamefont
  {Murugan}, \citenamefont {Thangadurai},\ and\ \citenamefont
  {Weppner}}]{doi:10.1002/anie.200701144}%
  \BibitemOpen
  \bibfield  {author} {\bibinfo {author} {\bibfnamefont {R.}~\bibnamefont
  {Murugan}}, \bibinfo {author} {\bibfnamefont {V.}~\bibnamefont
  {Thangadurai}}, \ and\ \bibinfo {author} {\bibfnamefont {W.}~\bibnamefont
  {Weppner}},\ }\href {\doibase 10.1002/anie.200701144} {\bibfield  {journal}
  {\bibinfo  {journal} {Angewandte Chemie International Edition}\ }\textbf
  {\bibinfo {volume} {46}},\ \bibinfo {pages} {7778} (\bibinfo {year}
  {2007})}\BibitemShut {NoStop}%
\bibitem [{\citenamefont {Meier}\ \emph {et~al.}(2014)\citenamefont {Meier},
  \citenamefont {Laino},\ and\ \citenamefont
  {Curioni}}]{doi:10.1021/jp5002463}%
  \BibitemOpen
  \bibfield  {author} {\bibinfo {author} {\bibfnamefont {K.}~\bibnamefont
  {Meier}}, \bibinfo {author} {\bibfnamefont {T.}~\bibnamefont {Laino}}, \ and\
  \bibinfo {author} {\bibfnamefont {A.}~\bibnamefont {Curioni}},\ }\href
  {\doibase 10.1021/jp5002463} {\bibfield  {journal} {\bibinfo  {journal} {The
  Journal of Physical Chemistry C}\ }\textbf {\bibinfo {volume} {118}},\
  \bibinfo {pages} {6668} (\bibinfo {year} {2014})}\BibitemShut {NoStop}%
\bibitem [{\citenamefont {Mottet}\ \emph {et~al.}(2019)\citenamefont {Mottet},
  \citenamefont {Marcolongo}, \citenamefont {Laino},\ and\ \citenamefont
  {Tavernelli}}]{PhysRevMaterials.3.035403}%
  \BibitemOpen
  \bibfield  {author} {\bibinfo {author} {\bibfnamefont {M.}~\bibnamefont
  {Mottet}}, \bibinfo {author} {\bibfnamefont {A.}~\bibnamefont {Marcolongo}},
  \bibinfo {author} {\bibfnamefont {T.}~\bibnamefont {Laino}}, \ and\ \bibinfo
  {author} {\bibfnamefont {I.}~\bibnamefont {Tavernelli}},\ }\href {\doibase
  10.1103/PhysRevMaterials.3.035403} {\bibfield  {journal} {\bibinfo  {journal}
  {Phys. Rev. Materials}\ }\textbf {\bibinfo {volume} {3}},\ \bibinfo {pages}
  {035403} (\bibinfo {year} {2019})}\BibitemShut {NoStop}%
\bibitem [{\citenamefont {Jalem}\ \emph {et~al.}(2013)\citenamefont {Jalem},
  \citenamefont {Yamamoto}, \citenamefont {Shiiba}, \citenamefont {Nakayama},
  \citenamefont {Munakata}, \citenamefont {Kasuga},\ and\ \citenamefont
  {Kanamura}}]{jalem_concerted_2013}%
  \BibitemOpen
  \bibfield  {author} {\bibinfo {author} {\bibfnamefont {R.}~\bibnamefont
  {Jalem}}, \bibinfo {author} {\bibfnamefont {Y.}~\bibnamefont {Yamamoto}},
  \bibinfo {author} {\bibfnamefont {H.}~\bibnamefont {Shiiba}}, \bibinfo
  {author} {\bibfnamefont {M.}~\bibnamefont {Nakayama}}, \bibinfo {author}
  {\bibfnamefont {H.}~\bibnamefont {Munakata}}, \bibinfo {author}
  {\bibfnamefont {T.}~\bibnamefont {Kasuga}}, \ and\ \bibinfo {author}
  {\bibfnamefont {K.}~\bibnamefont {Kanamura}},\ }\href {\doibase
  10.1021/cm303542x} {\bibfield  {journal} {\bibinfo  {journal} {Chemistry of
  Materials}\ }\textbf {\bibinfo {volume} {25}},\ \bibinfo {pages} {425}
  (\bibinfo {year} {2013})}\BibitemShut {NoStop}%
\bibitem [{\citenamefont {Boilot}\ \emph {et~al.}(1988)\citenamefont {Boilot},
  \citenamefont {Collin},\ and\ \citenamefont {Colomban}}]{BOILOT1988160}%
  \BibitemOpen
  \bibfield  {author} {\bibinfo {author} {\bibfnamefont {J.}~\bibnamefont
  {Boilot}}, \bibinfo {author} {\bibfnamefont {G.}~\bibnamefont {Collin}}, \
  and\ \bibinfo {author} {\bibfnamefont {P.}~\bibnamefont {Colomban}},\ }\href
  {\doibase https://doi.org/10.1016/0022-4596(88)90065-5} {\bibfield  {journal}
  {\bibinfo  {journal} {Journal of Solid State Chemistry}\ }\textbf {\bibinfo
  {volume} {73}},\ \bibinfo {pages} {160 } (\bibinfo {year}
  {1988})}\BibitemShut {NoStop}%
\bibitem [{\citenamefont {Ahmad}\ \emph {et~al.}(1987)\citenamefont {Ahmad},
  \citenamefont {Wheat}, \citenamefont {Kuriakose}, \citenamefont {Canaday},\
  and\ \citenamefont {Mcdonald}}]{NASICON_2}%
  \BibitemOpen
  \bibfield  {author} {\bibinfo {author} {\bibfnamefont {A.}~\bibnamefont
  {Ahmad}}, \bibinfo {author} {\bibfnamefont {T.}~\bibnamefont {Wheat}},
  \bibinfo {author} {\bibfnamefont {A.}~\bibnamefont {Kuriakose}}, \bibinfo
  {author} {\bibfnamefont {J.}~\bibnamefont {Canaday}}, \ and\ \bibinfo
  {author} {\bibfnamefont {A.}~\bibnamefont {Mcdonald}},\ }\href@noop {}
  {\bibfield  {journal} {\bibinfo  {journal} {Solid State Ionics}\ }\textbf
  {\bibinfo {volume} {24}},\ \bibinfo {pages} {89 } (\bibinfo {year}
  {1987})}\BibitemShut {NoStop}%
\bibitem [{\citenamefont {Hoover}(1985)}]{PhysRevA.31.1695}%
  \BibitemOpen
  \bibfield  {author} {\bibinfo {author} {\bibfnamefont {W.~G.}\ \bibnamefont
  {Hoover}},\ }\href {\doibase 10.1103/PhysRevA.31.1695} {\bibfield  {journal}
  {\bibinfo  {journal} {Phys. Rev. A}\ }\textbf {\bibinfo {volume} {31}},\
  \bibinfo {pages} {1695} (\bibinfo {year} {1985})}\BibitemShut {NoStop}%
\bibitem [{\citenamefont {Zipoli}\ and\ \citenamefont {Curioni}(2013)}]{fzi}%
  \BibitemOpen
  \bibfield  {author} {\bibinfo {author} {\bibfnamefont {F.}~\bibnamefont
  {Zipoli}}\ and\ \bibinfo {author} {\bibfnamefont {A.}~\bibnamefont
  {Curioni}},\ }\href {http://stacks.iop.org/1367-2630/15/i=12/a=123006}
  {\bibfield  {journal} {\bibinfo  {journal} {New Journal of Physics}\ }\textbf
  {\bibinfo {volume} {15}},\ \bibinfo {pages} {123006} (\bibinfo {year}
  {2013})}\BibitemShut {NoStop}%
\bibitem [{\citenamefont {Plimpton}(1995)}]{Plimpton1995}%
  \BibitemOpen
  \bibfield  {author} {\bibinfo {author} {\bibfnamefont {S.}~\bibnamefont
  {Plimpton}},\ }\href {\doibase 10.1006/JCPH.1995.1039} {\bibfield  {journal}
  {\bibinfo  {journal} {J. Comput. Phys.}\ }\textbf {\bibinfo {volume} {117}},\
  \bibinfo {pages} {1} (\bibinfo {year} {1995})}\BibitemShut {NoStop}%
\bibitem [{\citenamefont {Han}\ \emph {et~al.}(2018)\citenamefont {Han},
  \citenamefont {Linfeng}, \citenamefont {Jiequn},\ and\ \citenamefont
  {Weinan}}]{deepmd}%
  \BibitemOpen
  \bibfield  {author} {\bibinfo {author} {\bibfnamefont {W.}~\bibnamefont
  {Han}}, \bibinfo {author} {\bibfnamefont {Z.}~\bibnamefont {Linfeng}},
  \bibinfo {author} {\bibfnamefont {H.}~\bibnamefont {Jiequn}}, \ and\ \bibinfo
  {author} {\bibfnamefont {E.}~\bibnamefont {Weinan}},\ }\href {\doibase
  10.1006/JCPH.1995.1039} {\bibfield  {journal} {\bibinfo  {journal} {Computer
  Physics Communications}\ ,\ \bibinfo {pages} {178}} (\bibinfo {year}
  {2018})}\BibitemShut {NoStop}%
\bibitem [{\citenamefont {Kuhn}\ \emph {et~al.}(2013)\citenamefont {Kuhn},
  \citenamefont {Duppel},\ and\ \citenamefont {Lotsch}}]{C3EE41728J}%
  \BibitemOpen
  \bibfield  {author} {\bibinfo {author} {\bibfnamefont {A.}~\bibnamefont
  {Kuhn}}, \bibinfo {author} {\bibfnamefont {V.}~\bibnamefont {Duppel}}, \ and\
  \bibinfo {author} {\bibfnamefont {B.~V.}\ \bibnamefont {Lotsch}},\ }\href
  {\doibase 10.1039/C3EE41728J} {\bibfield  {journal} {\bibinfo  {journal}
  {Energy Environ. Sci.}\ }\textbf {\bibinfo {volume} {6}},\ \bibinfo {pages}
  {3548} (\bibinfo {year} {2013})}\BibitemShut {NoStop}%
\bibitem [{\citenamefont {Ong}\ \emph {et~al.}(2013)\citenamefont {Ong},
  \citenamefont {Mo}, \citenamefont {Richards}, \citenamefont {Miara},
  \citenamefont {Lee},\ and\ \citenamefont {Ceder}}]{ong_phase_2013}%
  \BibitemOpen
  \bibfield  {author} {\bibinfo {author} {\bibfnamefont {S.~P.}\ \bibnamefont
  {Ong}}, \bibinfo {author} {\bibfnamefont {Y.}~\bibnamefont {Mo}}, \bibinfo
  {author} {\bibfnamefont {W.~D.}\ \bibnamefont {Richards}}, \bibinfo {author}
  {\bibfnamefont {L.}~\bibnamefont {Miara}}, \bibinfo {author} {\bibfnamefont
  {H.~S.}\ \bibnamefont {Lee}}, \ and\ \bibinfo {author} {\bibfnamefont
  {G.}~\bibnamefont {Ceder}},\ }\href {\doibase 10.1039/C2EE23355J} {\bibfield
  {journal} {\bibinfo  {journal} {Energy \& Environmental Science}\ }\textbf
  {\bibinfo {volume} {6}},\ \bibinfo {pages} {148} (\bibinfo {year}
  {2013})}\BibitemShut {NoStop}%
\bibitem [{\citenamefont {Bhattacharya}\ and\ \citenamefont
  {Wolverton}(2016)}]{2016_Bhattacharya_Wolverton_Carbon_Sci_Tech}%
  \BibitemOpen
  \bibfield  {author} {\bibinfo {author} {\bibfnamefont {J.}~\bibnamefont
  {Bhattacharya}}\ and\ \bibinfo {author} {\bibfnamefont {C.~M.}\ \bibnamefont
  {Wolverton}},\ }\href@noop {} {\bibfield  {journal} {\bibinfo  {journal}
  {Carbon - Science and Technology}\ }\textbf {\bibinfo {volume} {8}},\
  \bibinfo {pages} {92} (\bibinfo {year} {2016})}\BibitemShut {NoStop}%
\bibitem [{\citenamefont {Kozinsky}\ \emph {et~al.}(2016)\citenamefont
  {Kozinsky}, \citenamefont {Akhade}, \citenamefont {Hirel}, \citenamefont
  {Hashibon}, \citenamefont {Els\"asser}, \citenamefont {Mehta}, \citenamefont
  {Logeat},\ and\ \citenamefont {Eisele}}]{kozinsky_effects_2016}%
  \BibitemOpen
  \bibfield  {author} {\bibinfo {author} {\bibfnamefont {B.}~\bibnamefont
  {Kozinsky}}, \bibinfo {author} {\bibfnamefont {S.~A.}\ \bibnamefont
  {Akhade}}, \bibinfo {author} {\bibfnamefont {P.}~\bibnamefont {Hirel}},
  \bibinfo {author} {\bibfnamefont {A.}~\bibnamefont {Hashibon}}, \bibinfo
  {author} {\bibfnamefont {C.}~\bibnamefont {Els\"asser}}, \bibinfo {author}
  {\bibfnamefont {P.}~\bibnamefont {Mehta}}, \bibinfo {author} {\bibfnamefont
  {A.}~\bibnamefont {Logeat}}, \ and\ \bibinfo {author} {\bibfnamefont
  {U.}~\bibnamefont {Eisele}},\ }\href {\doibase
  10.1103/PhysRevLett.116.055901} {\bibfield  {journal} {\bibinfo  {journal}
  {Physical Review Letters}\ }\textbf {\bibinfo {volume} {116}},\ \bibinfo
  {pages} {055901} (\bibinfo {year} {2016})}\BibitemShut {NoStop}%
\bibitem [{\citenamefont {Dai}\ \emph {et~al.}(2019)\citenamefont {Dai},
  \citenamefont {Chen}, \citenamefont {Glossmann},\ and\ \citenamefont
  {Lai}}]{DAI2019333}%
  \BibitemOpen
  \bibfield  {author} {\bibinfo {author} {\bibfnamefont {J.}~\bibnamefont
  {Dai}}, \bibinfo {author} {\bibfnamefont {Q.}~\bibnamefont {Chen}}, \bibinfo
  {author} {\bibfnamefont {T.}~\bibnamefont {Glossmann}}, \ and\ \bibinfo
  {author} {\bibfnamefont {W.}~\bibnamefont {Lai}},\ }\href {\doibase
  https://doi.org/10.1016/j.commatsci.2019.02.044} {\bibfield  {journal}
  {\bibinfo  {journal} {Computational Materials Science}\ }\textbf {\bibinfo
  {volume} {162}},\ \bibinfo {pages} {333 } (\bibinfo {year}
  {2019})}\BibitemShut {NoStop}%
\bibitem [{\citenamefont {Wang}\ and\ \citenamefont {Lai}(2015)}]{WANG2015612}%
  \BibitemOpen
  \bibfield  {author} {\bibinfo {author} {\bibfnamefont {Y.}~\bibnamefont
  {Wang}}\ and\ \bibinfo {author} {\bibfnamefont {W.}~\bibnamefont {Lai}},\
  }\href {\doibase https://doi.org/10.1016/j.jpowsour.2014.11.062} {\bibfield
  {journal} {\bibinfo  {journal} {Journal of Power Sources}\ }\textbf {\bibinfo
  {volume} {275}},\ \bibinfo {pages} {612 } (\bibinfo {year}
  {2015})}\BibitemShut {NoStop}%
\bibitem [{\citenamefont {Matsui}\ \emph {et~al.}(2014)\citenamefont {Matsui},
  \citenamefont {Takahashi}, \citenamefont {Sakamoto}, \citenamefont {Hirano},
  \citenamefont {Takeda}, \citenamefont {Yamamoto},\ and\ \citenamefont
  {Imanishi}}]{LLZO_exp}%
  \BibitemOpen
  \bibfield  {author} {\bibinfo {author} {\bibfnamefont {M.}~\bibnamefont
  {Matsui}}, \bibinfo {author} {\bibfnamefont {K.}~\bibnamefont {Takahashi}},
  \bibinfo {author} {\bibfnamefont {K.}~\bibnamefont {Sakamoto}}, \bibinfo
  {author} {\bibfnamefont {A.}~\bibnamefont {Hirano}}, \bibinfo {author}
  {\bibfnamefont {Y.}~\bibnamefont {Takeda}}, \bibinfo {author} {\bibfnamefont
  {O.}~\bibnamefont {Yamamoto}}, \ and\ \bibinfo {author} {\bibfnamefont
  {N.}~\bibnamefont {Imanishi}},\ }\href@noop {} {\bibfield  {journal}
  {\bibinfo  {journal} {Dalt. Trans.}\ ,\ \bibinfo {pages} {1019}} (\bibinfo
  {year} {2014})}\BibitemShut {NoStop}%
\bibitem [{\citenamefont {Perdew}\ \emph {et~al.}(1996)\citenamefont {Perdew},
  \citenamefont {Burke},\ and\ \citenamefont
  {Ernzerhof}}]{perdew_generalized_1996}%
  \BibitemOpen
  \bibfield  {author} {\bibinfo {author} {\bibfnamefont {J.~P.}\ \bibnamefont
  {Perdew}}, \bibinfo {author} {\bibfnamefont {K.}~\bibnamefont {Burke}}, \
  and\ \bibinfo {author} {\bibfnamefont {M.}~\bibnamefont {Ernzerhof}},\ }\href
  {\doibase 10.1103/PhysRevLett.77.3865} {\bibfield  {journal} {\bibinfo
  {journal} {Physical Review Letters}\ }\textbf {\bibinfo {volume} {77}},\
  \bibinfo {pages} {3865} (\bibinfo {year} {1996})}\BibitemShut {NoStop}%
\bibitem [{\citenamefont {Grimme}(2006)}]{grimme_semiempirical_2006}%
  \BibitemOpen
  \bibfield  {author} {\bibinfo {author} {\bibfnamefont {S.}~\bibnamefont
  {Grimme}},\ }\href {\doibase 10.1002/jcc.20495} {\bibfield  {journal}
  {\bibinfo  {journal} {Journal of Computational Chemistry}\ }\textbf {\bibinfo
  {volume} {27}},\ \bibinfo {pages} {1787} (\bibinfo {year} {2006})},\ \Eprint
  {http://arxiv.org/abs/https://onlinelibrary.wiley.com/doi/pdf/10.1002/jcc.20495}
  {https://onlinelibrary.wiley.com/doi/pdf/10.1002/jcc.20495} \BibitemShut
  {NoStop}%
\bibitem [{\citenamefont {Hutter}\ and\ \citenamefont {Iannuzzi}(2005)}]{CPMD}%
  \BibitemOpen
  \bibfield  {author} {\bibinfo {author} {\bibfnamefont {J.}~\bibnamefont
  {Hutter}}\ and\ \bibinfo {author} {\bibfnamefont {M.}~\bibnamefont
  {Iannuzzi}},\ }\href@noop {} {\bibfield  {journal} {\bibinfo  {journal}
  {Zeitschrift für Kristallographie}\ }\textbf {\bibinfo {volume} {220}},\
  \bibinfo {pages} {549} (\bibinfo {year} {2005})}\BibitemShut {NoStop}%
\bibitem [{\citenamefont {Belsky}\ \emph {et~al.}(2002)\citenamefont {Belsky},
  \citenamefont {Hellenbrandt}, \citenamefont {Karen},\ and\ \citenamefont
  {Luksch}}]{belsky_new_2002}%
  \BibitemOpen
  \bibfield  {author} {\bibinfo {author} {\bibfnamefont {A.}~\bibnamefont
  {Belsky}}, \bibinfo {author} {\bibfnamefont {M.}~\bibnamefont
  {Hellenbrandt}}, \bibinfo {author} {\bibfnamefont {V.~L.}\ \bibnamefont
  {Karen}}, \ and\ \bibinfo {author} {\bibfnamefont {P.}~\bibnamefont
  {Luksch}},\ }\href {\doibase 10.1107/S0108768102006948} {\bibfield  {journal}
  {\bibinfo  {journal} {Acta Crystallographica Section B Structural Science}\
  }\textbf {\bibinfo {volume} {58}},\ \bibinfo {pages} {364} (\bibinfo {year}
  {2002})}\BibitemShut {NoStop}%
\end{thebibliography}%
\end{document}